\documentclass[aps,prl,reprint,letterpaper,showpacs,amsmath,amssymb, superscriptaddress, twocolumn, preprintnumbers, prb]{revtex4-1}

\usepackage{graphicx}
\usepackage[absolute]{textpos}

\begin{document}
\begin{textblock*}{8.5in}(0.1in,0.25in)
\begin{center}
PHYSICAL REVIEW B \textbf{96}, 075160 (2017)
\end{center}
\end{textblock*}


\title{Electron spin dynamics of Ce$^{3+}$ ions in YAG crystals studied by pulse-EPR and pump-probe Faraday rotation}

\author{D. V. Azamat}
\affiliation{Institute of Physics of the Czech Academy of Sciences,
182 21 Prague 8, Czech Republic}
\author{V. V. Belykh}
\affiliation{Experimentelle Physik 2, Technische Universit\"{a}t Dortmund, D-44221 Dortmund, Germany}
\affiliation{P.N. Lebedev Physical Institute of the Russian Academy of Sciences, 119991 Moscow, Russia}
\author{D. R. Yakovlev}
\affiliation{Experimentelle Physik 2, Technische Universit\"{a}t Dortmund, D-44221 Dortmund, Germany}
\affiliation{Ioffe Institute, Russian Academy of Sciences, 194021 St. Petersburg, Russia}
\author{F. Fobbe}
\thanks{Present address: Photonics and Ultrafast Laser Science,
Ruhr Universit\"{a}t Bochum, D-44801 Bochum, Germany}
\affiliation{Experimentelle Physik 2, Technische Universit\"{a}t Dortmund, D-44221 Dortmund, Germany}
\author{D. H. Feng}
\affiliation{Experimentelle Physik 2, Technische Universit\"{a}t Dortmund, D-44221 Dortmund, Germany}
\affiliation{State Key Laboratory of Precision Spectroscopy, East China Normal University, Shanghai 200062, China}
\author{E. Evers}
\affiliation{Experimentelle Physik 2, Technische Universit\"{a}t Dortmund, D-44221 Dortmund, Germany}
\author{L.~Jastrabik}
\affiliation{Institute of Physics of the Czech Academy of Sciences,
182 21 Prague 8, Czech Republic}
\author{A. Dejneka}
\affiliation{Institute of Physics of the Czech Academy of Sciences,
182 21 Prague 8, Czech Republic}
\author{M. Bayer}
\affiliation{Experimentelle Physik 2, Technische Universit\"{a}t Dortmund, D-44221 Dortmund, Germany}
\affiliation{Ioffe Institute, Russian Academy of Sciences, 194021 St. Petersburg, Russia}

\received{27 July 2017} \published{30 August 2017}
%

\begin{abstract}
The spin relaxation dynamics of Ce$^{3+}$ ions in heavily cerium-doped YAG crystals is studied using pulse-electron paramagnetic resonance and time-resolved pump-probe Faraday rotation. Both techniques address the 4$f$ ground state, while pump-probe Faraday rotation provides also access to the excited 5$d$ state. We measure a millisecond spin-lattice relaxation time $T_1$, a microsecond spin coherence time $T_2$ and a $\sim 10$~ns inhomogeneous spin dephasing time $T_2^*$ for the Ce$^{3+}$ ground state at low temperatures.
The spin-lattice relaxation of Ce$^{3+}$ ions is due to modified Raman processes involving the optical phonon mode at $\sim$ 125 cm$^{-1}$. The relaxation at higher temperature goes through a first excited level of the $^{2}$F$_{5/2}$ term at about $\hbar \omega \approx 228$~cm$^{-1}$. Effects provided by the hyperfine interaction of the Ce$^{3+}$  with the $^{27}$Al nuclei are observed.
\\
\doi{10.1103/PhysRevB.96.075160}
\end{abstract}

\maketitle

\section{Introduction}
Cerium-doped yttrium-aluminum garnet (YAG:Ce) materials have attracted attention due to their potential application in high performance scintillators developed earlier for lasers. Further, the triply-ionized cerium ion in YAG is considered to be a suitable system for quantum bit implementations \cite{Kolesov, Siyushev, Lewis, Reyher, Ronda, Romanov, Badalyan,Tolmachev2017,Vedda2002,Asatryan2014}.

Reyher et al.\cite{Reyher} identified the Ce$^{3+}$ ion as origin of the photo-refractive effect in YAG crystals by using a combination of optically-detected magnetic resonance and magnetic
circular dichroism of the absorption. They showed a correlation between the optical bands and the $4f \leftrightarrow 5d$ transitions of the Ce$^{3+}$ states.

The effective $g$ factors of the Ce$^{3+}$ centers measured by electron paramagnetic resonance (EPR) \cite{Lewis} provide an accurate assessment of the $^{2}$F$_{5/2}$ ground state that results from mixing of the three Kramers doublets by the spin-orbit coupling and orthorhombic crystal field interactions. In the present work we confirm that according to an analysis of the temperature dependence of the EPR linewidth, the first excited level of the $^{2}$F$_{5/2}$ multiplet is about 228~cm$^{-1}$ above the ground level. The necessity to consider the influence of defects on the relaxation processes affecting these states was investigated by Aminov et al. \cite{Aminov1,Aminov2,Aminov3}.

EPR is a technique suitable for investigating the spin dynamics of the $4f$ ground state of Ce$^{3+}$. The spin dynamics of the excited $5d$ state at room temperature was investigated by detection of the optically-induced magnetization \cite{Kolesov2007} and pump-probe Faraday rotation \cite{Liang2017}. All these measurements were performed on ensembles of Ce ions. Recently, spin control and manipulation of a single Ce$^{3+}$ ion was demonstrated at room and low temperatures using time-resolved fluorescence technique \cite{Kolesov,Siyushev}.

In this paper we study YAG:Ce crystals heavily doped at 0.1 at. \% and 0.5 at. \% of Ce$^{3+}$. Our goal is to perform a comparative study of the electron spin dynamics by two experimental techniques, pulse-EPR and pump-probe Faraday rotation, in order to obtain comprehensive information about the longitudinal and transverse spin relaxation times and the different underlying spin relaxation processes of the ground and excited states of the Ce$^{3+}$ centers.
We demonstrate the possibility to coherently manipulate the coupled electron-nuclear states of the Ce and Al ions in the YAG matrix. The coherent spin dynamics of the excited 5$d$ state is detected by pump-probe Faraday rotation up to high temperatures of about 200~K, while for the ground $4f$ state it diminishes above $\sim 40$~K.

\section{Experimental details}

Two commercial samples of YAG:Ce$^{3+}$ crystals with  Ce$^{3+}$ ion concentrations of 0.1 at. \% and 0.5 at. \% and 0.5~mm thickness are studied. The samples were fabricated by the Hangzhou Shalom Electro-optics Technology Co., Ltd.

The magnetic resonance studies are performed using a X-band Bruker FT-EPR ELEXSYS 580 spectrometer, operating at a microwave frequency of 9.7~GHz. The temperature is varied in a helium cryostat from 2.5
to 300~K. The Electron-Spin-Echo (ESE) detected EPR spectra are taken from monitoring the intensity of the Hahn echo as function of magnetic field. The spin-lattice relaxation rate of Ce$^{3+}$ is measured by using the inversion recovery pulse sequence ($\pi-t-\pi/2-\tau-\pi-\tau-$inverted echo) \cite{Jeschke}. The lower limit of the relaxation rate measurements is approximately 100~ns.

In the pump-probe Faraday rotation measurements the samples are placed in a helium cryostat with a split-coil vector magnet, which allows us to apply an external magnetic field at an arbitrary angle relative to the sample without the need for sample rotation. The sample temperature is varied from 2 up to 200~K. We use a pulsed Ti:Sapphire laser operating at 906~nm wavelength which is frequency doubled with a BBO crystal to a wavelength of 453~nm. The laser emits pulses with 2~ps duration at a repetition rate of 76~MHz (repetition period $T_\mathrm{R}=13$~ns). The laser beam is split into a circularly polarized pump beam, which generates the electron spin polarization, and a linearly polarized probe beam whose Faraday rotation is measured after transmission through the sample. To probe the temporal evolution of the spin polarization, the delay between the pump and probe pulses is varied within a 6~ns interval using a mechanical delay line. Also the resonant spin amplification technique \cite{Kikkawa1998,Yugova2012,Glazov2008} is used for measuring spin dephasing times exceeding $T_\mathrm{R}$.

\section{Results}

\subsection{Magnetic resonance of YAG:Ce$^{3+}$}

\begin{figure}
\includegraphics[width=0.8\columnwidth]{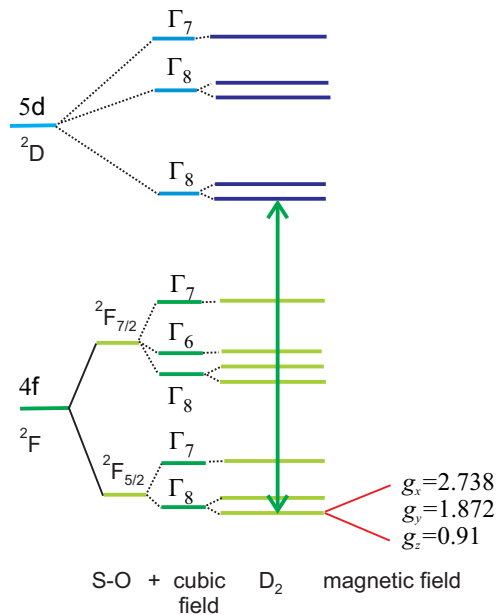}
\caption{Combined action of the crystal field and the spin-orbit interaction on the 4$f$ and 5$d$ levels of Ce$^{3+}$.}
\label{fig:Ce}
\end{figure}

The triply-ionized cerium ions in YAG occupy $c$-sites with local symmetry D$_{2}$. The details of YAG crystal structure can be found in Refs. \cite{Gurin2015,Kostic2015}. The structure of spin-orbital multiplets is schematically shown in Fig.~\ref{fig:Ce}. The $^{2}$F(4\textit{f}$^{1}$) multiplet is split into the $^{2}$F$_{5/2}$ and $^{2}$F$_{7/2}$ terms by the combined action of the spin-orbit coupling and the crystal field. The ground state doublet is separated by 228~cm$^{-1}$ from the higher lying doublet of the $^{2}$F$_{5/2}$ ($\Gamma_8$) term \cite{Kolesov}. In an external magnetic field its splitting is strongly anisotropic, see below the parameters for Eq.~\eqref{eq:1}.  Light absorption populates the 5$d$ excited state (see arrow in Fig.~\ref{fig:Ce}), which has a roughly isotropic $g$-factor tensor with $g=2$.

The X-band (9.78~GHz) EPR spectra of the ground doublet of $^{2}$F$_{5/2}$ term can be described by using an orthorhombic spin Hamiltonian \cite{Lewis} (with  effective spin $S$ = 1/2):
\begin{eqnarray} \label{eq:1}
H = \mu_B g_z B_z S_z + \mu_B g_xB_x S_x + \mu_B g_y B_y S_y ,
\end{eqnarray}
where the $g_i$ are the $g$ tensor components: $g_x=2.738$, $g_y=1.872$, $g_z=0.91$. The $z$, $x$ and $y$
principal axes of the $g$ tensor are parallel to the [110], [001], and [1$\bar{1}$0] crystal directions, respectively. The magnetic multiplicity of the Ce$^{3+}$ EPR spectra is $K_M=6$.

\begin{figure}
\includegraphics[width=0.9\columnwidth]{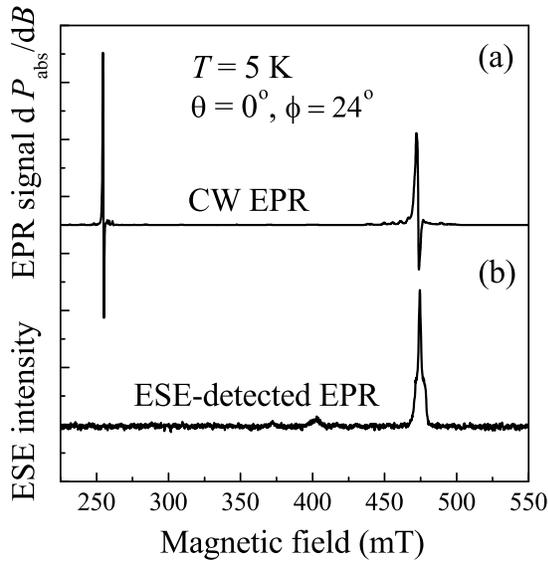}
\caption{Comparison of continuous wave (CW) EPR spectrum (a) with ESE-detected field-swept EPR spectrum (b) of the Ce$^{3+}$ centers in YAG:Ce (0.5 at. \%). The magnetic field orientation is about $\mathbf{B}\|$[001].}
\label{fig:CWvsESE}
\end{figure}

Figure~\ref{fig:CWvsESE}a shows a continuous wave (CW) EPR spectrum of the Ce$^{3+}$ centers ($T= 5$~K) in an applied magnetic field defined by the polar angles of $\theta$ = 0$^{\circ}$, $\phi$ = 24$^{\circ}$.
Thus, the applied field $\mathbf{B}$ is oriented along the [001] direction. $dP_{\mathrm{abs}}/dB$ is the derivative of the microwave absorption. The double line at 255~mT in the EPR spectrum is contributed by two magnetically-nonequivalent centers. A quadruple line occurs at about 474.5~mT. In Fig.~\ref{fig:CWvsESE}(b) the ESE detected EPR spectrum is shown for the same field orientation. Contrary to the CW-EPR spectrum it shows only the high-field line. The low-field resonance does not show up due to short phase relaxation time $T_2$.

\begin{figure}
\includegraphics[width=0.9\columnwidth]{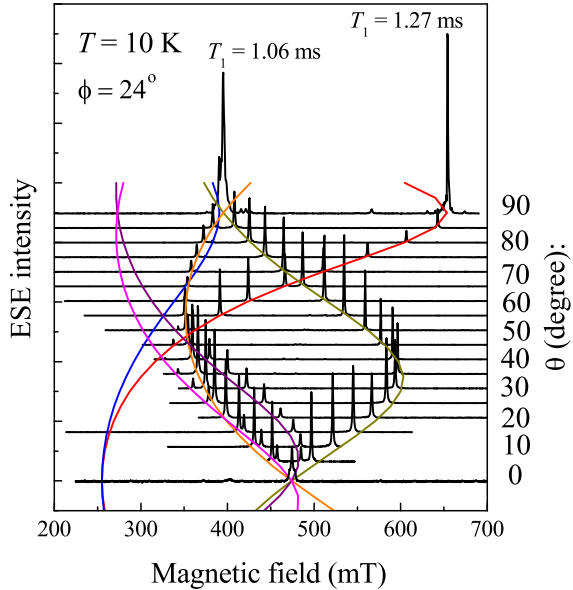}
\caption{Waterfall-plot of ESE-detected EPR spectra for the cerium centers in YAG:Ce (0.5 at. \%). The applied magnetic field is swept for different $\theta$ angles, $\phi=24^{\circ}$. The EPR magnetic multiplicity is $K_M=6$. The spin-lattice relaxation times $T_1$ are indicated for the low- and high-field lines for $\theta = 90^{\circ}$. Solid lines show the calculated positions of the EPR resonances after Eq.~\eqref{eq:1}.}
\label{fig:EPRAnisotr}
\end{figure}

Figure~\ref{fig:EPRAnisotr} presents the angle variation of the  Ce$^{3+}$ spectra measured by the spin-echo method. The complex pattern of EPR resonances is due to six magnetically nonequivalent (and crystallographically equivalent) Ce$^{3+}$ sites with D$_2$ point symmetry in the YAG unit cell. For comparison the colored lines show the calculated positions of resonances via angle $\theta$. The resonance lines disappear in the low magnetic field range due to fast spin-spin relaxation. The different intensities of the lines near the same magnetic field in the Ce$^{3+}$ ESE-EPR spectra reflect different occupations between crystallographically equivalent $c$-site positions. These occupation preferences are especially  visible when the applied magnetic field is oriented close to the main directions of $\theta=0^{\circ}$ or $90^{\circ}$. E.g. at $\theta=10^{\circ}$ four lines from different $c$-cite positions are visible in the ESE-EPR spectrum at $B = 450-500$~mT, but their intensities are not equal. Such facet-related site selectivity was previously investigated for the Neodym-ions in YAG: Nd$^{3+}$ \cite{Wolfe}.

\begin{figure}
\includegraphics[width=0.9\columnwidth]{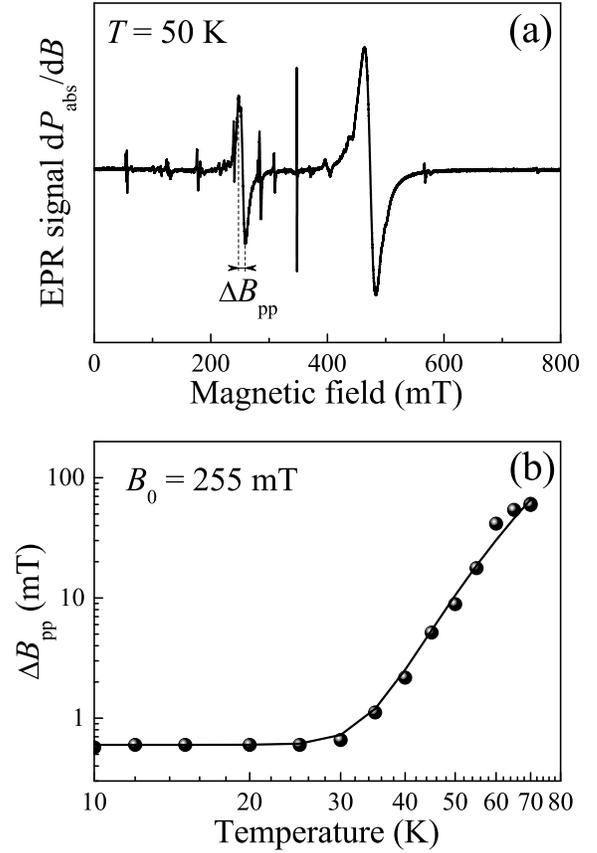}
\caption{(a) CW-EPR spectrum of Ce$^{3+}$ in YAG:Ce (0.5 at.\%) for $\textbf{B} \|$[001] ($\theta=0^{\circ}$) and $T=50$~K.
(b) Temperature dependence of peak-to-peak line-width of the low-field line ($B_0=255$~mT). Solid
line is fit to experimental data by Eq.~\eqref{eq:Orb} with the
the excited state that becomes thermally populated at $\sim 228$~cm$^{-1}$.}
\label{fig:Linewidth}
\end{figure}

Our studies demonstrate some peculiarities in the spin-lattice relaxation of the Ce$^{3+}$ ions. Figure~\ref{fig:Linewidth}(a) shows that the CW-EPR resonance lines broaden with increasing temperature up to 50~K due to rapid spin-lattice relaxation above 25~K. The temperature dependence of the
peak-to-peak linewidth $\Delta B_{pp}$ for the Ce$^{3+}$ line at $B_0 = 255$~mT is shown in Fig.~\ref{fig:Linewidth}(b). The linewidth is constant at $\Delta B_{pp}=0.6$~mT for $T<25$~K, but increases for higher temperatures reaching 60~mT at 70~K.  The experimental data can be fit by the equation:
\begin{equation}
\label{eq:Orb}
\Delta B_{pp}=A + \frac {C} {\exp(\hbar \omega / k_{B}T)-1}.
\end{equation}
where $A=0.6$~mT, $C=7000$~mT and  $\hbar \omega \approx 228$~cm$^{-1}$. The second term here is provided by the relaxation involving optical vibrations. The energy of $\hbar \omega$ is approximately equal to the energy splitting to the next higher lying Kramers doublet (the upper-lying $^2$F$_{5/2}$ ($\Gamma_8$) level in Fig.~\ref{fig:Ce}). Note that there is also LO phonon mode in YAG with energy of 227~cm$^{-1}$ \cite{Hurrell}.

\begin{figure}
\includegraphics[width=0.9\columnwidth]{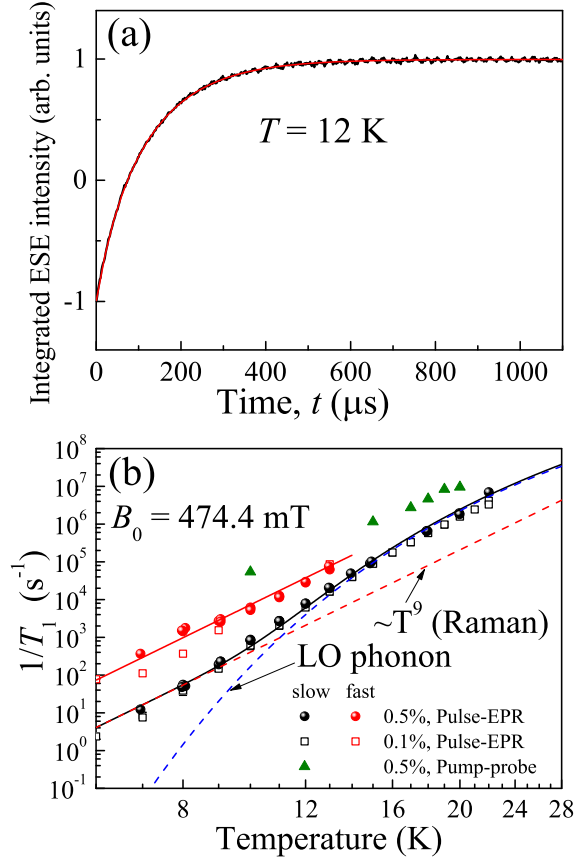}
\caption{(a) Three-pulse echo decay (black line) for YAG:Ce (0.5 at.\%), with $\mathbf{B}\|$[001] ($\theta=0^{\circ}$), $B_0=474.4$~mT  and $T = 12$~K. Red line is a bi-exponential fit to the experimental data. (b) Temperature dependence of the spin-lattice relaxation rates $1/T_1$ corresponding to the two
components in the magnetization kinetics. For comparison, the experimental data for YAG:Ce (0.5 at.\%) are shown by solid circles and for YAG:Ce (0.1 at.\%) by open squares. Slow and fast components of the dynamics are given by black and red colors, respectively. Green triangles show data measured by pump-probe Faraday rotation for YAG:Ce (0.5 at.\%). Solid lines are fits with Eq.~\eqref{eq:RaLO}. Dashed lines show the contributions to the relaxation due to the two-phonon Raman and optical phonon modified processes.} \label{fig:T1EPR}
\end{figure}

In addition, the spin-lattice relaxation for the Ce$^{3+}$ ions in YAG is investigated in the low temperature range using the inversion recovery pulse sequence ($\pi-t-\pi/2-\tau-\pi-\tau-$inverted echo) \cite{Jeschke}. Figure ~\ref{fig:T1EPR}(a) shows the inversion recovery curve measured at $T=12$~K and $B_0=474.4$~mT. The magnetization kinetics is fitted using a bi-exponential function with characteristic relaxation times of $T_{1,i}$=36~$\mu$s and 133~$\mu$s.

The temperature dependence of the relaxation rates for heavily doped YAG:Ce (0.5 and 0.1 at. \%) samples are mainly determined by a two-phonon Raman ($\thicksim T^9$) process modified by the presence of a longitudinal optical (LO) phonon mode (solid lines in Fig.~\ref{fig:T1EPR}(b)):
\begin{equation}
\label{eq:RaLO}
\frac{1}{T_{1,i}} = A_i T^9 + B_i \exp(-\frac{\Delta E}{k_\text{B} T}),
\end{equation}
where $A_i$ and $B_i$ are constants, the index $i=1, 2$ accounts for the fast and slow components in the bi-exponential recovery curve, and $\Delta E$ is the LO phonon energy. We obtain an acceptable
fit with $A_1=2.4 \cdot 10^{-7}$~s$^{-1}$K$^{-9}$, $B_1=1.4 \cdot 10^{10}$~s$^{-1}$, $A_2=2.4 \cdot 10^{-6}$~s$^{-1}$K$^{-9}$, $B_2=8 \cdot 10^{10}$~s$^{-1}$ and $\Delta E = 125$~cm$^{-1}$. Raman
studies in YAG have revealed LO phonons with an energy of approximately 125~cm$^{-1}$ \cite{Hurrell}. Thus, we see that the  spin-lattice relaxation of Ce$^{3+}$ ions in YAG:Ce (0.5 and 0.1 at. \%) is occurring through a two-phonon Raman process modified by interactions with optical vibrations.

Figure~\ref{fig:EPRAnisotr} illustrates that the relaxation times become smaller in the low field region: \textit{T}$_{1} = 1.06$~ms for the low field line and \textit{T}$_{1}= 1.27$~ms for high field line ($\theta=90^{\circ}$, $T=10$~K). The measured slight orientational dependence of the relaxation rate at the X-band microwave frequency is, supposedly, due to the low crystal field symmetry of the Ce$^{3+}$ ions in a YAG lattice.

\begin{figure}
\includegraphics[width=0.9\columnwidth]{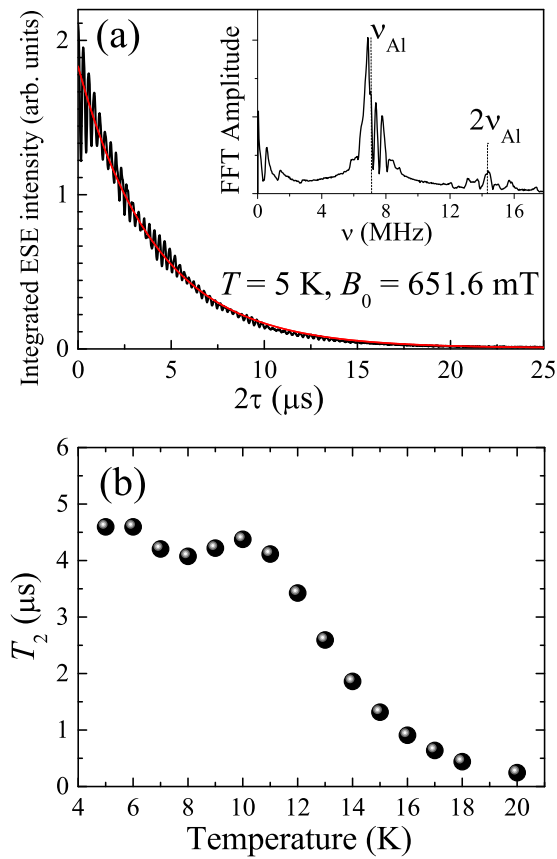}
\caption{(a) Two-pulse decay plot for Ce in YAG:Ce (0.5 at. \%) at $T = 5$~K. The experimental Hahn echo decay is shown {\sl vs} the delay time $2\tau$. Red line is fit with the phase relaxation time $T_2=4.5$~$\mu$s. The inset shows the fast Fourier transform of the ESE nuclear modulation.
(b) Phase relaxation time $T_2$ as function of temperature. The measurements are done for the high-field transition at $B_0=651.6$~mT, the magnetic field orientation $\theta=90^{\circ}$, $\phi=24^{\circ}$.}
\label{fig:Echo}
\end{figure}

The experiments show an anomalous character of the phase relaxation \textit{T}$_{2}$ of Ce$^{3+}$ in YAG. Figure~\ref{fig:Echo}(a) presents the two-pulse Hahn echo decay for Ce in YAG:Ce (0.5 at. \%) with phase memory time $T_2=4.5$~$\mu$s at $T=5$~K. The high field resonance line was measured in the orientation of applied magnetic field $\theta=90^{\circ}$. Note that in the $\theta=0^{\circ}$ orienatation the ESE nuclear modulation makes it more complicated to determine the phase relaxation time.

Figure~\ref{fig:Echo}(b) illustrates changes of the phase memory time $T_2$ with increasing temperature. Around 7~K the relaxation time displays a local minimum and then a steep decrease for $T>12$~K. Generally, the temperature dependence of the $T_{2}$ time does not follow that for $T_{1}$. The more rapid ESE decay
can be explained by cross-relaxation processes involving spin-spin flips of resonant Ce$^{3+}$ ions. Considering the influence of the surrounding aluminum nuclei, the decay of the ESE signal involves flip-flop mechanism with the $^{27}$Al nuclear spins \cite{Salikhov}.

The modulation of the Hahn echo signal in Fig.~\ref{fig:Echo}(a) results from the coherent precession of the coupled electron-nuclear spins, which evidences the interaction between the unpaired electron of Ce$^{3+}$ and the surrounding $^{27}$Al nuclei. The aluminum nuclei which possess a nonzero electric quadrupole moment were extensively used to probe the local charge distribution in YAG crystals \cite{Brog,Edmonds,Charnaya,Badalyan}. The effective Hamiltonian for analyzing the interaction with the $^{27}$Al nuclei ($\textit{I}$ = 5/2) can be written in the form:
\begin{eqnarray} \label{eq:Hamilt}
H_{eff} = \sum_{i}(\mathbf{S} \cdot \mathbf{A}_{i} \cdot
\mathbf{I}_{i} + \mathbf{I}_{i} \cdot \mathbf{Q}_{i} \cdot
\mathbf{I}_{i} - g_{N} \mu_{N} \mathbf{B} \cdot \mathbf{I}_{i})
\end{eqnarray}
where $\mathbf{S}$ is the electron spin operator,  $\mathbf{Q}_{i}$ is a traceless matrix, $\mathbf{A}_{i}$ the hyperfine interaction tensor,  $\mathbf{I}_{i}$ nuclear spin operator of interaction with $i$-th aluminum nucleus, $\mu_{N}$ the nuclear magneton, $g_{N}$ the nuclear $g$ factor of $^{27}$Al and the applied magnetic field orientation is $\theta$ = 90$^{\circ}$. Summation on $i$ is for three magnetically nonequivalent positions of Al. In the case of axial symmetry the quadrupole coupling \textit{P} = 3/2$\textit{Q}_{zz}$. The quadrupole interaction parameter can be written as \textit{P} =3\textit{e}$^{2}$\textit{q}\textit{Q}/[4\textit{I}(2\textit{I} - 1)], where $q$ is the gradient of the electrical field, and $Q$ is the nuclear quadrupole moment. The Fourier transform of the ESE modulated spectrum shows the nuclear Larmor frequency $\nu_{\mathrm{Al}}= 7.235$~MHz (resonance line of $B_0=651.6$~mT) and a set of five nuclear quadrupole split  lines ($\textit{I}$ = 5/2).  The quadrupole split spectrum reflects the axial symmetry of the  electric field gradient about the [001]-type directions and corresponds to the interaction with aluminum ions in \textit{d}-sites.  For the tetrahedral $\textit{d}$-site of aluminum the quadrupole splitting value is $P/h = 0.35$~MHz in the principal $z$, $x$, $y$ axes parallel to the [001], [100] and [010] directions, respectively. It is equal to the energy separation between adjacent lines in the quadrupole multiplet (i.e. five nuclear resonance lines, I = 5/2). The quadrupole coupling constant for tetrahedral $d$-sites in YAG:Ce (0.5 at. \%) is $e^2qQ/h = 4.67$~MHz.

\begin{figure}
\includegraphics[width=0.9\columnwidth]{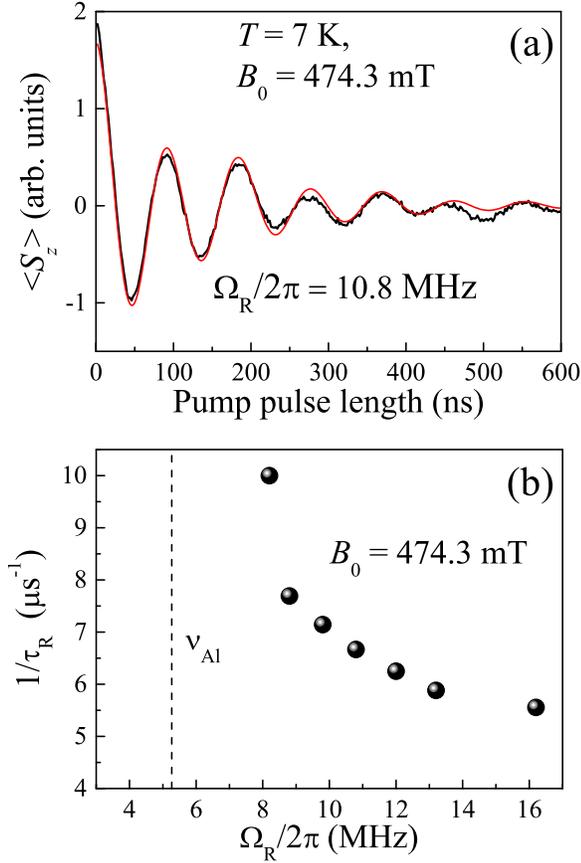}
\caption{(a) Rabi oscillations: dependence of the inverted echo signal on the pump pulse length (black line). Red line is fit to experimental data using  Eq.~\eqref{eq:rabi} which gives $\tau_{R}=150$~ns.
(b) Experimental data of decay rate of Rabi oscillations {\sl vs} Rabi frequency. Damping rate of Rabi oscillations increases in vicinity of aluminum nuclear Larmor frequency of $\nu_{Al} = 5.266$~MHz shown by dashed line. (a),(b) Magnetic field $\mathbf{B}\|$[001] ($\theta=0^{\circ}$), $B_0=474.4$~mT.}
\label{fig:Rabi}
\end{figure}

We investigate the Rabi oscillations taken on the Ce$^{3+}$ resonance using the inverted spin-echo signal at the applied field $\mathbf{B}\|$[001]. In order to describe the echo dynamics it is convenient to consider the spins whose Rabi frequencies are distributed around the $^{27}$Al nuclear Larmor frequency. The nutation experiments are performed with a pump pulse followed by a $\pi/2-\pi$ sequence after a delay time $\tau \sim T_{2}$. Figure~\ref{fig:Rabi}(a) shows the Rabi oscillation of Ce$^{3+}$ measured by varying the length of microwave pump pulse. The damping time of the Rabi oscillations, $\tau_R$, can be approximately evaluated by using the equation \cite{Jeschke,Baibekov}:
\begin{equation} \label{eq:rabi}
\langle S_z \rangle(t) = S_z \mid_{t=0}\exp({-t/\tau_R}) \cos(\Omega_R t) ,
\end{equation}
where $\Omega_R$ is the Rabi frequency. Fit shown by red line in Fig.~\ref{fig:Rabi}(a) allows us to determine $\Omega_R/2 \pi=10.8$~MHz and $\tau_R=150$~ns. Performing such measurements and fits for different microwave powers we receive the dependence of the decay rate of Rabi oscillations on the Rabi frequency (Fig.~\ref{fig:Rabi}(b)). One can see that the damping rate increases near the frequency range of the aluminum Larmor precession. The electron-nuclear cross-relaxation transitions in the rotating reference frame become more efficient at the $^{27}$Al nuclear Larmor frequency of $\nu_{Al} = 5.266$~MHz for $B_0=474.3$~mT. Fig.~\ref{fig:Rabi}(b) demonstrates that the decoherence of the Rabi oscillations is accompanied by energy dissipation into the aluminum nuclei spin bath. The damping time of the oscillations $\tau_{R}$ is much shorter in comparison to the phase relaxation time $T_{2}=1.35$~$\mu$s at $T=7$~K in $B_0=474.3$~mT and $\theta=0^{\circ}$.

\subsection{Pump-probe Faraday rotation spectroscopy}

In this section we discuss the Ce$^{3+}$ electron spin dynamics investigated by pump-probe Faraday rotation. The time-resolved pump-probe Faraday rotation is an established technique, which is widely used for investigation of coherent spin dynamics in solids, see e.g. Ref.~\cite{Ch6}. In this technique electron spin polarization is induced by circularly-polarized pump pulses and the polarization dynamics is detected via the Faraday rotation (FR) of linearly-polarized probe pulses, which are delayed in time relative to the pump pulses. The time resolution of this technique is limited to the laser pulse duration, which is 2~ps in our experiment. I.e., it is a few orders of magnitude better than in EPR which is limited to 100~ns.

The laser wavelength of the pump and probe beams was set to 453~nm, near the absorption maximum of the phonon-assisted transition between the lowest energy sublevels of the 4$f$ and 5$d$ multiplets. The $\sigma^+$ circularly polarized pump excites the electron from the 4$f$ (spin-down) state to the 5$d$ (spin-up) state. Due to spin relaxation of the 5$d$ electrons, their relaxation back to the 4$f$ states does not compensate the pump-induced spin polarization of the 4$f$ states. In this way, an ensemble of the Ce$^{3+}$ ions acquires a predominant spin orientation (spin-up) of the ground 4$f$ state \cite{Kolesov,Siyushev}. Therefore, in pump-probe experiments we get access to the electron spin dynamics of the 4$f$ state (which is usually investigated by EPR), but also to that of the 5$d$ state within its lifetime of about 65~ns \cite{Zych2000}. Note that due to the fast phonon-assisted relaxation within 4$f$ and 5$d$ multiplets we only consider their lower sublevels.

\subsubsection{Spin precession in transversal magnetic field}

\begin{figure}
\includegraphics[width=1\columnwidth]{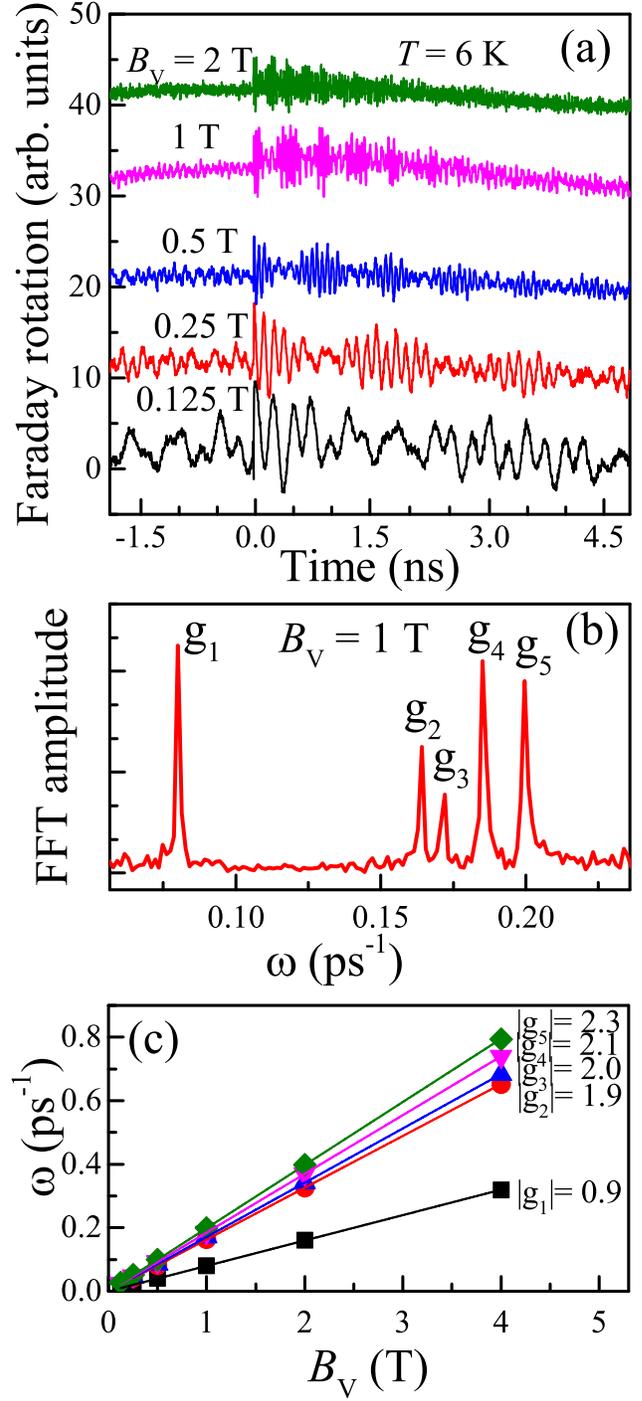}
\caption{(a) Dynamics of the Faraday rotation signal at different magnetic field strengths in Voigt geometry for the YAG:Ce (0.1 at. \%) sample at $T = 6$~K. (b) Fast Fourier transform (FFT) spectrum taken at $B_\text{V} = 1$~T for positive time delays. (c) Magnetic field dependences of the Larmor frequencies. Lines are linear fits to the data.}
\label{fig:PPKin}
\end{figure}

Figure~\ref{fig:PPKin}(a) shows the dynamics of the Faraday rotation signal of YAG:Ce (0.1 at. \%) in magnetic fields $B_\text{V}$ up to 2~T applied in the Voigt geometry, i.e., perpendicular to the light wave vector. The oscillating FR signal is composed of five Larmor precession frequencies, as evidenced by the fast Fourier transformation (FFT) spectrum in Fig.~\ref{fig:PPKin}(b). Figure~\ref{fig:PPKin}(c) shows the linear increase of each Larmor frequency with magnetic field. The five effective $g$ factors evaluated from the relation $\hbar \omega =|g| \mu_\mathrm{B} B_\text{V}$ fall in the range from 0.9 to 2.3. As we will show below they can be assigned to the ground $4f$ ($|g_{1}| \approx 0.9$, $|g_{2}| \approx 1.9$, $|g_{4}| \approx 2.1$, $|g_{5}| \approx 2.3$) and the excited $5d$ ($|g_{3}| \approx 2.0$) states. For each of this states, in general, one would expect six different frequencies corresponding to the different magnetically nonequivalent positions of Ce$^{3+}$ centers in YAG lattice, which characterized by the different orientations of the $g$ tensor.

\begin{figure}
\includegraphics[width=1\columnwidth]{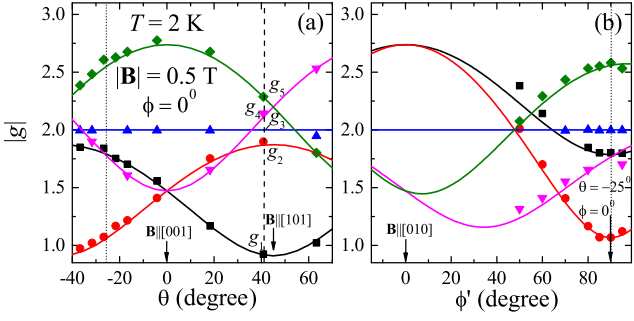}
\caption{Angular variation of Ce$^{3+}$ spin precession frequencies described by  effective $g$ factors. (a) The magnetic field vector $\mathbf{B}$ is rotated in the sample plane, corresponding to $\phi = 0^o$ (b) $\mathbf{B}$ is rotated out of the sample plane from the direction [010] (sample normal) to $\theta = -25^o$, $\phi = 0^o$ (sample plane) parametrized with the angle $\phi'$.
Solid lines show the calculated dependencies after Eq.~\eqref{eq:1}.  Dashed line shows the orientation corresponding to the results presented in Figs.~\ref{fig:PPKin} and~\ref{fig:RSA}. Dotted lines show the same orientation of $\mathbf{B}$ on both figures. The data are shown by symbols for the YAG:Ce (0.5 at. \%) sample at $T=2$~K.}
\label{fig:PPAnisotr}
\end{figure}

\begin{figure}
\includegraphics[width=0.8\columnwidth]{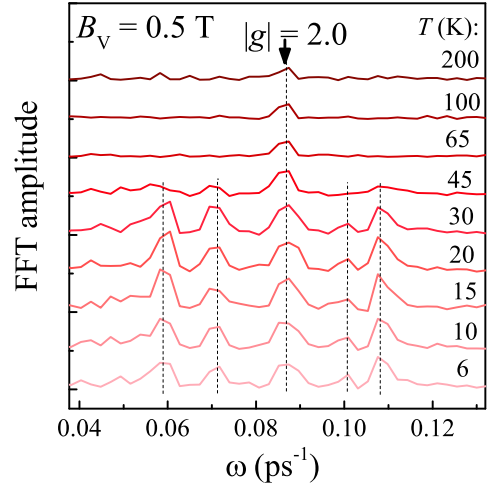}
\caption{Temperature dependence of Fast Fourier transform
(FFT) spectra from Faraday rotation signals taken at
different temperatures for $B_\text{V}=0.5$~T in Voigt geometry. The data are shown for YAG:Ce (0.1 at.\%).}
\label{fig:FFTTDep}
\end{figure}

Because of the low (rhombic) symmetry of the $c$-sites occupied by of Ce$^{3+}$ ions the Larmor frequencies show large variation for the  rotation of the magnetic field vector \textbf{B}. Figure~\ref{fig:PPAnisotr}(a) shows the $g$-factor variations when the field is rotated in the sample plane, corresponding to the variation of the angle $\theta$ between [001] and [100] axes at $\phi = 0^o$. Figure~\ref{fig:PPAnisotr}(b) shows the anisotropy, when the field is rotated out of the sample plane between the directions of [010] axis (sample normal) and $\theta = -25^o$, $\phi = 0^o$ (sample plane). The curves are fitted with the same $g$-tensor parameters as the positions of the EPR resonances (Fig.~\ref{fig:EPRAnisotr}). The only exception is the precession mode with an almost isotropic $g$ factor $\lvert g \rvert = 2.0$ (the blue line in Fig.~\ref{fig:PPAnisotr}), which we assign to the excited $5d$ state. The assignment of this mode is confirmed by the FFT spectra of the FR signals measured at different temperatures up to 200~K (Fig.~\ref{fig:FFTTDep}). All modes, except the one with $\lvert g \rvert =2.0$ vanishes for $T>45$~K, which allows us to assign them to the $4f$ state with a strong spin-orbit coupling that provides a shortening of the spin dephasing times at elevated temperatures. But the mode with $\lvert g \rvert =2.0$ is well resolvable up to $T=200$~K due to the comparatively much weaker spin orbit interaction of the $5d$ excited state. Recently, even its observation at room temperature was reported \cite{Liang2017}.

\begin{figure}
\includegraphics[width=0.8\columnwidth]{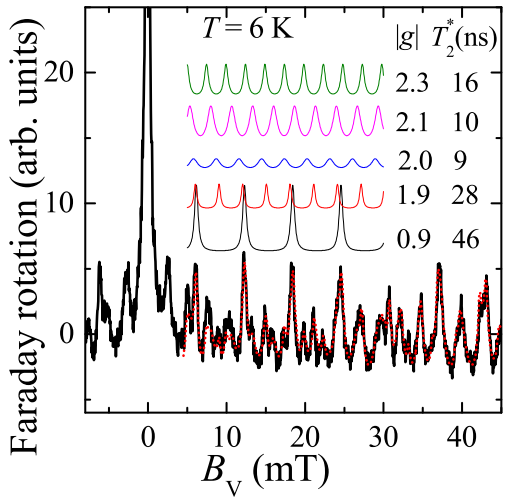}
\caption{Resonant spin amplification (RSA) signal (black line) measured for YAG:Ce (0.1 at. \%) at $T=6$~K. Dotted red line is a fit with Eq.~\eqref{eqn:RSA}, which enables us to determine the $g$ factors
and spin dephasing times $T_2^*$. The separate contributions associated with the different Larmor frequencies are shown above the experimental curve.}
\label{fig:RSA}
\end{figure}

In Fig.~\ref{fig:PPKin}(a), at negative time delays one sees oscillations originating from the preceding pump pulses. This shows that the decay time $T_2^*$ of the electron spin ensemble is comparable to or even exceeding the pulse repetition period $T_\mathrm{R}=13$~ns. To determine the $T_2^*$ at $B\approx 0$ we perform resonant spin amplification (RSA) experiments \cite{Kikkawa1998} by scanning the magnetic field applied in Voigt geometry and monitoring the Faraday rotation signal at a small negative delay $\Delta t \approx -30$~ps (Fig.~\ref{fig:RSA}). The RSA curve shows peaks corresponding to integer numbers of spin oscillations between subsequent laser pulses. The widths of the peaks allow one to determine the $T_2^*$ of the
precession modes through fitting the RSA curve to the following spin polarization equation:\cite{Glazov2008, Yugova2012}
\begin{align} S(B_\text{V})=\sum_{i=1}^5 \frac{S_i}{2} \times
\frac{\cos(g_i \mu_\mathrm{B} B_\text{V} T_\mathrm{R}) -e^{-T_\mathrm{R}/{T}_{\mathrm{2},i}^*}}
{\cosh(T_\mathrm{R}/{T}_{\mathrm{2},i}^*)-\cos(g_i\mu_\mathrm{B} B_\text{V} T_\mathrm{R})}, \label{eqn:RSA}
\end{align}
where we take into account $|\Delta t| \ll T_2^*, T_\text{R}$.
This equation involves the sum of the five components. Each of them describes the RSA curve for a single oscillation mode with amplitude $S_i$ (spin polarization created by one pump pulse), $g$ factor $g_i$  and decay time ${T}_{2,i}^*$. The $g$ factors determined from the fit are in agreement with those evaluated in Fig.~\ref{fig:PPKin}(c). The obtained $T_2^*$ values are in the range from 9 to 46~ns and are listed in Fig.~\ref{fig:RSA}. The variation of $T_2^*$ for the ground state precession frequencies originates from the low crystal field symmetry position of the cerium ions in the YAG lattice.

\subsubsection{Longitudinal spin relaxation}

\begin{figure}
\includegraphics[width=0.8\columnwidth]{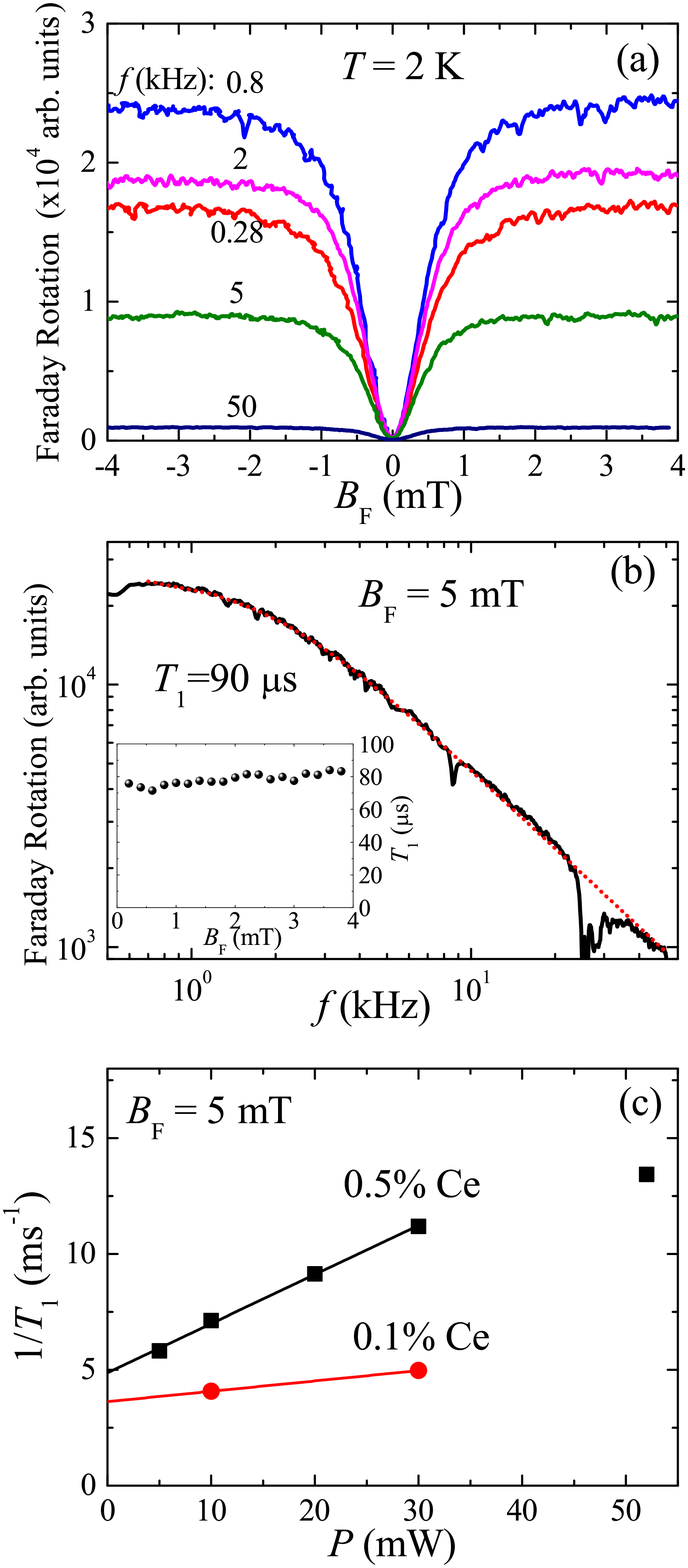}
\caption{(Color online)(a) Polarization recovery curves for various pump modulation frequencies.
(b) Frequency dependence of the Faraday rotation signal at $B_\text{F}=5$~mT. The spikes in the signal are caused by the hardware. Dotted line is fit with Eq.~\eqref{fdep}. Inset shows magnetic field dependence of $T_1$.
(c) Pump power dependence of spin-lattice relaxation rate $1/T_1$ for YAG samples with different Ce$^{3+}$ concentrations.
Data in panels (a) and (b) are shown for YAG:Ce (0.5 at. \%), $P=30$~mW. In (a)-(c) $T=2$~K.}
\label{fig:fDep}
\end{figure}

In order to study the longitudinal spin relaxation characterized by the $T_1$ time, we measure the polarization recovery curves for different modulation frequencies $f$ by switching the circular polarization of the pump between $\sigma^+$ and $\sigma^-$ \cite{Heisterkamp2015}. The magnetic field, $B_\text{F}$, was applied in the Faraday geometry parallel to the light wave vector. The magnetic field dependence of the Faraday rotation signal was measured at a small negative delay between the pump and probe pulses [Fig.~\ref{fig:fDep}(a)]. At zero external field the generated spin polarization precesses about randomly oriented local magnetic fields, e.g., provided by the nuclear spins, which reduces the spin polarization component parallel to the pump wave vector that we detect via the Faraday rotation. In magnetic fields $B_\text{F}$ exceeding the local fields the spin polarization is stabilized along the external field direction, which increases the FR amplitude.
This characteristic behavior can be seen in Fig.~\ref{fig:fDep}(a), where the FR amplitude strongly increases and becomes saturated for $B_\text{F}>1$~mT. The half width at half maximum (HWHM) of the signal increase, $\Delta B\approx0.6$~mT, characterizes the strength of the local magnetic fields. Note that similar local field values of $0.4-0.6$~mT were observed in optically detected magnetic resonance experiments in Ref.~\onlinecite{Siyushev} (after recalculation from the width of ODMR spectrum resonance) and attributed to the exchange fields of the nuclei spins of neighboring Al ions. Indeed, as we have shown in the previous section, the electron spin dynamics of the $4f$ state is strongly influenced by the interaction with $^{27}$Al nuclei (Fig.~\ref{fig:Echo}).

According to the model of Ref.~\cite{Merkulov2002}, for an isotropic electron $g$ factor the decrease of the electron polarization at $B=0$ due to the frozen nuclear field should go down to $1/3$ of the value at higher $B_\text{F}$. The fact that in our case the signal drops almost to zero might indicate that either the nuclear magnetic field is oriented always in the sample plane, which is unlikely, or that the nuclear spins are not frozen on the timescale of the longitudinal electron spin relaxation time $T_1$, which can take place if the electron spin interacts with only one or a few nuclear spins.

Let us consider that only a small fraction of Ce$^{3+}$ centers is subject to spin orientation, i.e. we operate in the linear regime of the impact of the laser excitation density. In case of $T_1 > T_\mathrm{R}$ and fixed circular polarization of the pump, spin polarization will accumulate under pulsed periodic excitation \cite{Salis2001, Belykh2016a}. In this case the Faraday rotation signal is proportional to $S_0 T_1/T_\text{R}$, where $S_0$ is the amplitude of the signal created by one laser pulse. Note, that $S_0$ is close to the amplitude of the spin precession, when $\bf{B}$ is applied in the Voigt geometry [Fig.~\ref{fig:PPKin}(a)].

In our experiment we modulate the pump circular polarization at frequency $f$. In case $1/f < T_1$, the signal is no longer determined by $T_1/T_\text{R}$, but by the modulation frequency. Indeed, the periodic switching of the pump polarization from $\sigma^+$ to $\sigma^-$ prevents the accumulation of electron spin polarization. As a result, the Faraday rotation signal decreases with increasing $f$ [Fig.~\ref{fig:fDep}(b)]. This allows us to evaluate the $T_1$ time by the spin inertia method \cite{Heisterkamp2015}. The following dependence of the Faraday rotation amplitude on $f$ is expected:
\begin{equation}
S(f)=\frac{S_0 T_1 /T_\text{R}}{\sqrt{1+(2\pi f T_1)^2}}.
\label{fdep}
\end{equation}
Using this relation for fitting the data in Fig.~\ref{fig:fDep}(b) we evaluate $T_1=0.09$~ms at $B_\text{F}=5$~mT. One can see in the inset of Fig.~\ref{fig:fDep}(b) that $T_1$ time evaluated by the spin inertia method is slowly increasing with $B_\text{F}$ in the range of low magnetic fields.

The $T_1$ extracted from the fit depends on the pump power $P$, presumably due to the saturation effect and sample overheating. At low $P$ the dependence of $1/T_1$ on $P$ is almost linear [Fig.~\ref{fig:fDep}(c)], allowing one to extrapolate the dependence to $P=0$, where $T_1=0.28$~ms for YAG:Ce (0.1 at. \%) and $T_1=0.21$~ms for 0.5 at. \%.  In Ref.~\cite{Siyushev} $T_1\approx 4.5$~ms was measured for single Ce$^{3+}$ ions at $T = 3.5$~K and $B=49$~mT. Keeping in mind that the $5d$ state lifetime is only 65~ns, we conclude that the measured sub-millisecond $T_1$ times correspond to the $4f$ ground state.

\begin{figure}
\includegraphics[width=0.9\columnwidth]{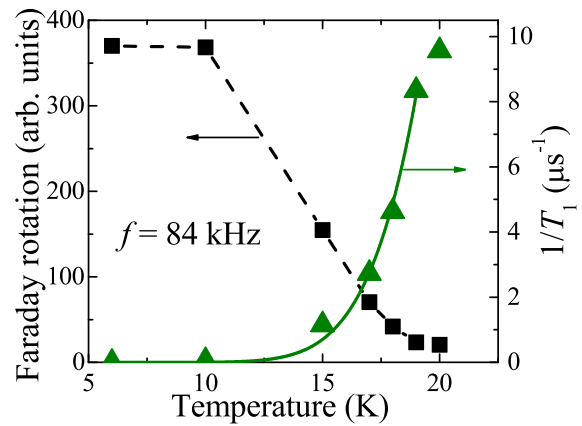}
\caption{Temperature dependence of Faraday rotation signal at $B_\text{F}\approx3$~mT. Left axis gives
Faraday rotation signal shown by squares, and right axis gives spin-lattice relaxation rate $1/T_1$ shown by triangles. Data are given for YAG:Ce (0.5 at. \%) at $f=84$~kHz modulation frequency. Dashed line is guide for the eye. Solid line is fit to the data using Eq.~\eqref{eq:RaLO}.}
\label{fig:PPT1TDep}
\end{figure}

The Faraday rotation signal at $|B|\gtrsim 1$~mT decreases with increasing temperature in a threshold-like way (see the squares in Fig.~\ref{fig:PPT1TDep}). At low temperatures the ratio $T_1 \gg 1/f$ is fulfilled for $f = 84$~kHz and Eq.~\eqref{fdep} reduces to $S (T = 0) \approx S_0 / (2\pi f T_\text{R})$.
Using this relation with Eq.~\eqref{fdep} we obtain $1/T_1(T) = 2\pi f \sqrt{S^2(T = 0)/S^2(T)-1}$, which can be used for evaluation of $T_1$ at $T > 10$~K. The temperature dependence of $1/T_1$ derived in this way (the triangles in Fig.~\ref{fig:PPT1TDep}) is fitted with Eq.~\eqref{eq:RaLO}, where we set $A=2.4 \cdot 10^{-6}$~s$^{-1}$K$^{-9}$. $B \approx 9 \times 10^{10}$~s$^{-1}$ and $\Delta E \approx 15$~meV (120 cm$^{-1}$) are derived from this fit. These parameters are in good agreement with those determined from the dependence of $1/T_1$ on temperature in the previous section [Fig.~\ref{fig:T1EPR}(b)], confirming the dominant phonon relaxation mechanism in the studied temperature range. Some discrepancy between the $T_1$ values obtained by the two methods [Fig.~\ref{fig:T1EPR}(b)] increasing at lower temperatures, is related to the fact that the pump-probe experiments were done at pump power $P \approx 20$~mW where $T_1$ is underestimated $\sim 2$ times [Fig.~\ref{fig:fDep}(c)]. This effectively corresponds to the somewhat larger lattice temperature (by $1-2$~K). The other possible source of the discrepancy is magnetic field dependence of $T_1$.


\section{Conclusions}

The electron spin dynamics of Ce$^{3+}$ ions in a YAG crystal has been studied by pulse-EPR and pump-probe Faraday rotation. The spin relaxation times measured by both techniques are in reasonable agreement with each other. We have shown show that the techniques are complementary: both of them provide information on the 4$f$ ground states, while the pump-probe Faraday rotation give also access to the short living 5$d$ excited states. The measured times for the Ce$^{3+}$ ground state at low temperatures, are milliseconds for the spin-lattice relaxation time $T_1$, microseconds for the spin coherence time $T_2$, and $\sim 10$~ns for the inhomogeneous spin dephasing time $T_2^*$. The underlying relaxation mechanisms have been discussed. It was shown that at low temperature range the spin-lattice relaxation is governed by the optical phonon mode ($\sim 125$~cm$^{-1}$) modified Raman processes, while at higher temperatures the spin relaxation dominates by two-stage process through the higher lying Kramers doublet separated by $\hbar \omega \approx 228$~cm$^{-1}$. Nutation experiments revealed the strong damping of the Rabi oscillations decay when Rabi frequency is nearby to the $^{27}$Al nuclear Larmor frequency. Finally, we have also found effects provided by the hyperfine interaction of the Ce$^{3+}$  with $^{27}$Al nuclei. The nuclear quadrupole coupling constant for tetrahedral aluminum $d$-sites in YAG:Ce (0.5 at. \%) lattice obtained in the current study is $e^{2} qQ/h = 4.67$~MHz.

$\mathbf{Acknowledgments}$
We are grateful to A.~Greilich and S.~B.~Orlinskii for valuable advices and useful discussions, and to E.~Kirstein and E.~A.~Zhukov for help with experiments. The EPR experiments had financial support of the Ministry of Education, Youth and Sport of the Czech Republic (project No.  LO1409) and the Czech Science Foundation – GACR (project No. 16-22092S). Pump-probe experiments were supported by the Deutsche Forschungsgemeinschaft in the frame of ICRC TRR 160 and by the Russian Science Foundation (Grant No. 14-42-00015).

\end{document}